\documentclass[11pt,a4paper]{article}
\pdfoutput=1
\usepackage[hidelinks]{hyperref}
\usepackage{doi}
\usepackage[margin=1.4cm]{geometry}
\usepackage[utf8]{inputenc}
\usepackage[T1]{fontenc}
\usepackage{graphicx}
\usepackage{multicol}
\usepackage{changepage}
\usepackage{caption}
\usepackage{textcomp}
\usepackage{cite}
\usepackage{sectsty}

\sectionfont{\fontsize{12}{12}\selectfont}

\newcommand\blfootnote[1]{
    \begingroup
    \renewcommand\thefootnote{}\footnote{#1}
    \addtocounter{footnote}{-1}
    \endgroup
}

\begin{document}
\begin{center}
\textbf{\Large{On the alternatives to the ideal mathematical points-like separatedness}}
\vspace{0.22cm}

Bartosz Jura \\
\blfootnote{E-mail address: barbartekjura@gmail.com}
\vspace{0.05cm}
\begin{adjustwidth}{2cm}{2cm}
	In a recent paper as an alternative to models based on the notion of ideal mathematical point, characterized by a property of separatedness, we considered a viewpoint based on the notion of continuous change, making use of elements of a non-classical logic, in particular the fuzzy sets theory, with events represented as spatiotemporally blurred blobs. Here we point out and discuss a number of aspects of this imperfect symbolic description that might potentially be misleading. Besides that, we analyze its relation to various concepts used commonly to model physical systems, denoted by terms like: point, set, continuous, discrete, infinite, or local, clarifying further how our viewpoint is different and asking whether, in light of our main postulate, any of these notions, or their opposites, if exist, are in their usual meanings suitable to accurately describe the natural phenomena. \\
\end{adjustwidth}
\vspace{0.4cm}

\end{center}
\begin{multicols}{2}
\section{Introduction}

Recall the main claim considered in \cite{Jura2024} that in the natural world there is no fact of the matter about the spatio-temporal separation of events, or, in other words, about their separatedness\footnote{Notion not related (at least not directly) to the one from algebraic geometry with the same name. Not related also to the separation axioms for topological spaces \cite{sepax}, to the notion of spaces' \cite{spacesep} or sets' \cite{setsep} separability, nor other notions with names alluding to the commonsense intuition of 'separation' between (or \textit{of}) things, each having its very own specific technical definition.}, meaning that it is not adequate to represent and model real world events (or, states) as ideal mathematical points, as current approaches typically tend to do. As an alternative, we considered instead a viewpoint based on the notion of continuous change (CC), using elements of a non-classical logic, in particular the fuzzy sets theory, with events represented as (spatiotemporally) blurred blobs. Is this symbolic description clear, readily comprehensible, and unambiguous, or rather prone to misinterpretations and potentially misleading? Even if it is the former, is it also clear how exactly it differs from the separatedness-based framework\footnote{Alternatively, the term 'paradigm' could be used here, although 'framework' seems to us more adequate and illustrative, as describing something to which a whole construction is fit and by which it is supported. Primarily for this reason this latter term is chosen as the one that will be used throughout this paper.} (SF), especially since we adopted and use in different contexts and with altered meaning some common terms which, when used within the SF, have their specific definitions (like, among others, change that is claimed to be \textit{continuous}, or the use of fuzzy \textit{sets})? If so, is it more accurate and thus better than the SF, that is, does it allow to describe (we avoid using the term \textit{explain} here, reserving it for specific physical theories, certain general aspects of which, for ones finding support in empirical data, we may wish to recover or capture within our viewpoint) the empirical observations and findings in a more natural (meaning: simpler, more 'intuitive', and also more comprehensive) way (and maybe to make some new testable predictions as well)? Also, is it the only possible alternative view, arising uniquely from negating (or generalizing) the SF?

To answer these questions, we need to know first of all what exactly is the main aspect of the SF that we claim is deficient (in the context of its ambitions of being a basis for accurate descriptions of natural phenomena) and we are trying to find an alternative to. Since we insist that it relates with the notion of mathematical point being its central element, what exactly is to be understood then by the 'point' that we keep refering to? Mathematical point is typically considered a primitive notion being an abstract idealization of an exact position (in physical space, or time, or in an abstract mathematical space) \cite{geo}. It is considered zero-dimensional. It thus has no shape or size. It is conventional, however, to draw a filled round shape to represent it, as in Figure \ref{Fig1}a. Is this an accurate and/or appropriate representation? Is it a harmless practice to draw it like this?

\section{What is the shape of a mathematical point}

Although it is just a convention, which is not to be taken overly seriously, nonetheless in our view it at least contributes (together with the supposition about its size, as discussed further below) to two interrelated effects, that is, it: 1) diverts attention from the essence of the SF's central notion, and 2) suggests that possible to represent and model accurately within the SF alone might be phenomena which are in fact not.

The essence of the SF's central notion, a point (which might represent events, states, or other abstract objects), is in our view that it is supposed to represent something that is distinct from other points. Primarily and directly, when drawn symbolically on a white piece of paper (as in Figure \ref{Fig1}), it is distinct from the white space (which might represent an 'empty' space, a 'gap') that surrounds it. This is the separatedness property: either something is one and the same thing, or not the same and thus separate, distinct things (as depicted for example in ref. \cite{Jura2022}'s Figure 3A). In the SF, a point is simply something different, and (which is particularly salient point) it can be different only along a given specific dimension\footnote{Note that there are many different definitions of 'dimension' \cite{dim}. For example, a topological (covering) dimension of a discrete space of isolated points is considered equal 0. Here, by the number of dimensions we will mean, somewhat more informally, the minimum number of coordinates needed to specify any point within a space, with the number of dimensions of a discrete set of points when drawn on a piece of paper that is assumed to be 2-dimensional being equal 2, and with the coordinates of such points defined possibly as the number of points along a given dimension needed to reach a given point when 'jumping' between 'neighbouring' points starting from some baseline level. And by dimension itself, we will mean one of the directions in which one can 'move' between the points.} (e.g., along an $x$-axis, with some $x_1$ being distinct from $x_2$). When drawn on a plane that is assumed to be exactly two-dimensional (like the piece of paper), it is something different along each of the dimensions being represented (in this example along one of the two, that is, vertically or horizontally), and thus more faithfully to the intent of the SF (and less deceptively) a point should be depicted as a rectangular shape, as in Figure \ref{Fig1}b\footnote{'Moving' (i.e., when comparing points) along the diagonal, and out of the corner, as one could possibly try here, is not really an option, as what is supposed to be represented here are two distinct dimensions (there is thus no 'intermediate' dimension). In fact, when the points represent actual things, this is not possible due to there being no exact (sub-)point that we consider here a 'corner' (of the point), as we shall discuss in yet more detail later.}. The popular round shape is misleading as it 'distracts' from this essence of a point as being simply distinct from others, and, besides that, it suggests that in the SF one could represent accurately a round shape, that in a sense changes 'smoothly', in not any specific dimension (or, practically speaking, in more than one dimension at a time, as if it was gradually 'merging' together the dimensions).

In fact, within the SF (that is, using only the elements, or 'building blocks', that it offers and permits), it is not possible to represent precisely (that is, with a precision that would be exact and not merely an 'arbitrary' one) a curve, which can be considered as something changing smoothly, in a sense in more than one direction at once (or, more precisely speaking, in no specific direction), as the one in Figure \ref{Fig1}c, left. If we assume two dimensions, we are allowed then to take steps from one point onto another either along horizontal or vertical dimension, but not along the two at the same time (Figure \ref{Fig1}c, right). Unless we were rotating the coordinate axes by a certain degree in incremental steps, only then, at each such incremental step, we could 'move' along a new dimension, which would be at a certain angle to the preceding one, 'visiting' different points along this new dimension. The variable representing the angle, however, relative to a baseline level, would itself form a dimension of the same kind, that is, with the angle value represented at each step again as a distinct separate point along the $angle$-axis. To draw a 'smooth' curve (in the informal sense of a shape with no abrupt, sharp turns) as in Figure \ref{Fig1}c, left, is actually to just not respect the strict rules of the game as imposed (on itself) by the SF, trying to prove that possible might be something which in fact is not. One draws what is assumed to be a smooth curve (which never goes exactly vertically, horizontally, or in any other specific direction), and then tries to describe it somehow using exact numbers (represented here by the points' locations), while they are not applicable as unable to represent a smooth, gradual change\footnote{One could suggest, as one of conceivable arguments in defence of SF in this respect, that the curve depicted in Figure \ref{Fig1}c, left, represents in fact a point in a space with a curvilinear coordinate axis, and it is distinct (from other points) along one specific (although intrinsically curved) dimension, and such curvilinear system of reference can be defined abstractly, using exact numbers, rebutting thus our claim of the SF's capabilities being overrated. Our response to this would be that the point is that what is not possible to represent precisely within the SF is not (an exact value of) an abstract notion of curvature, which obviously it is, but rather an actual change (note that here, describing change, we do not mean 'actual' to be synonymous with 'physical'), where a given magnitude never takes any specific value, be it angle, or curvature, or anything else. In other words, this example with the change of angle (with the curve possibly, but not necessarily, representing some physical motion) is just one specific instance from the general, abstract class of phenomena in application to which the SF seems to be helpless (albeit undoubtedly resourceful, always providing one with means to make up to a certain extent for this deficiency, effectively masking its true limits.).}. At least part of this illusion (of, at least, a potential possibility of attaining the exact precision) can be attributed to this assumption (or, convention), saying that any of the points, being parts of which the curve supposedly consists, actually has no shape, or can be taken to be of arbitrary shape, especially round, and thus, as the argument goes, from such points surely we could construct a 'smooth' curvy shape. Another assumption, likely also contributing here, as we have already mentioned, is that those parts have no (or at most 'negligible') size.

\section{What is the size of a mathematical point}

Does it make any sense, however, to represent, and try to 'reconstruct', any object, and especially any real, physically existing thing, with parts which have no size? While it is far from obvious what is to be understood by "physically existing", or by "size", let us assume for now that something with no size, that is, occupying literally no part of a relevant space (physical space, or time, or some other abstract mathematical space), simply does not exist and then there is nothing to talk about. It is assumed in the SF that continuous\footnote{Note that there is also a variety of distinct concepts with names refering to the notion of something being 'continuous' \cite{cont,Felscher2000}. Here, when refering to the SF, by 'continuous' we will mean, again perhaps rather informally, a property of a collection of objects any subset of which can be put in one-to-one correspondence with a real interval, describing the fact that however short of an interval we took we will always find that it consists of (relatively) even shorter separate pieces.}, that is, infinitely divisible intervals, consist of (only) such points with no size (we remark and stress here that CC is not divisible and thus not continuous in this sense). We posit that, in the SF itself, by virtue of being something distinct from others (and, primarily, from the surrounding 'gap'), any point (assumed therein to be 'zero-dimensional') will in fact have a non-zero size, or in other words, an 'extension' (which from the outside, that is, relative to others, might be arbitrarily small, but evident once the picture is 'zoomed in', as visualized in ref. \cite{Jura2022}'s Figure 3A).

Deliberately and explicitely an 'extended' version of ideal point-like objects seems to be used in certain models being described commonly as 'discrete', refering to a version of discreteness that is supposed to describe what we could consider a 'physical continuum', that is, a situation where different entities (possibly including events, defined as a spacetime points) can be considered in some sense (not necessarily amenable to precise definition) to be 'situated directly next to each other'. This kind of discreteness means then that an interval (containing some objects) is not infinitely divisible, but consists instead of some 'minimal-size' indivisible units (as in the top two rows in Figure \ref{Fig1}d), with their size expressed (implicitly) in terms of the number of some 'underlying' points of certain fixed size (with the lower bounds on size possibly defined separately, independently for \begin{center}
\fbox{\includegraphics[width=0.76\linewidth,keepaspectratio]{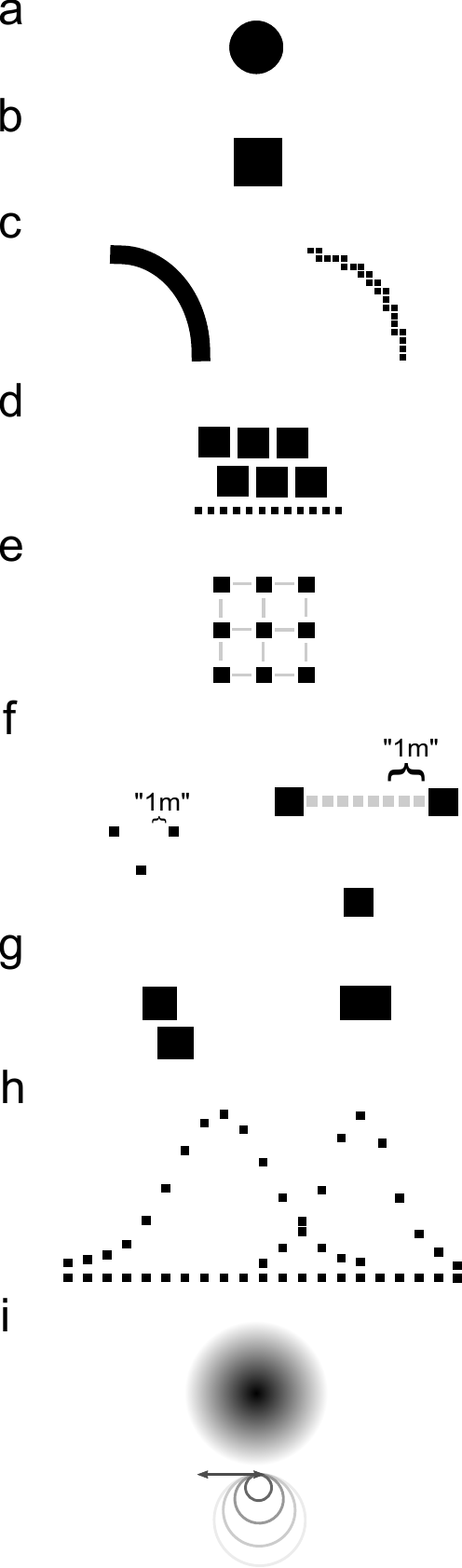}}
\captionof{figure}{The ideal mathematical points-like separatedness and an alternative. The panels' contents are discussed and explained throughout the main text. \label{Fig1}}
\end{center}different such 'units'). Hovewer, what if what is about to be (conceptually) divided is actually embedded (in some way) in an 'underlying' background (as represented by the bottom row of points, for visual clarity drawn below the other two, in Figure \ref{Fig1}d). Then, how to choose the origin of the division? For a given origin of partition chosen (as one of the top two rows in Figure \ref{Fig1}d), some points of the underlying structure can fall in between and others within the interiors of some of the extended points, with the latter suggesting that 'their' respective extended points could be divided further, as being embedded in the domain which is 'more' divisible, while the former being left in the 'void', not 'assigned' to any of the extended points (with this 'split' into two 'kinds' of underlying points depending on the partition; one could add also a third kind, of 'undecided' points, that is ones partly within an extended point and partly within the void). Perhaps such an indivisibility would be more justifiable without such underlying background, when the points to be (potentially) divided are the only thing there is, like in an abstract graph, as in Figure \ref{Fig1}e. However, in an abstract graph like this, beside the links (depicted here as the gray rectangular points) there remain still the 'gaps' in between the nodes and the links, with the links themselves being just like points, with all the ensuing issues.

Actually, in the context of this particular kind of discreteness, it would seem to us more appropriate to call such extended points (being like, for example, the ancient notion of atoms, as bits of matter, with minimal assumed sizes) continuous rather than discrete. For when an individual point is considered in itself, from the 'inside', and not as embedded in some underlying at least potentially more divisible background, then what stands out is its continuity, that is, the 'connectedness' (or, 'persistence') of the stuff that it represents (in our examples, the black ink used to draw a point). It is rather in this sense that we call change 'continuous'. CC would be thus more akin to a discrete rather than continuous interval in this regard, however definitely not in the sense of discretely spaced, isolated points (which are also assumed to have no size) as discussed next.

Another version of discreteness\footnote{This one being more directly related to what is formally, in the abstract, considered a discrete space \cite{disc}. Again, we hope that we sufficiently faithfully represent how the various mathematics-related terms and concepts appear to be typically used in different models aimed at describing the physical systems, however since the issue and question of how exactly a given concept, as used in some practically-oriented model, is related to its formal counterpart, and whether it shares all its particular properties or consequences, is often left partially or entirely unanswered, maybe for future investigations, as currently too complex or perhaps irrelevant (for instance, can an actual, i.e., physical, solid object, in three-dimensional space considered continuous, be decomposed into two exact copies, in terms of size, of the original one? \cite{tar}), for formal definitions and treatment interested reader is refered to one of the strictly mathematics-oriented resources as referenced below or to other, relevant, literature.} (which, hovewer, may result from the one considered above) might be used to describe a situation where a given quantity can take only some selected, isolated values, and not others. For all purposes involving the object or process considered (that is one being described by a given quantity), this kind of discreteness implies that in between the isolated permitted points (each representing a distinct value) there is an empty space which cannot be occupied, which however can equivalently be considered as being filled with a kind of 'invisible', 'unattainable', but extended point of certain size (with any interval of the isolated points corresponding possibly one-to-one to an interval of natural numbers; in this particular kind of applications such model might describe, for example, an oscillation that could occur with only some particular values of frequency but not others, possibly 'jumping', in some way, between the different, spaced values, but not a given oscillation with different frequencies actually situated next to itself, whatever that would mean, as in the 'continuum' of different actual entities as described above, which is an important distinction to keep in mind). The isolated points themselves are then assumed to have no size, just like in the continuous case in this regard. As far as representing natural phenomena with ideal points goes, the essence of our arguments, as outlined thus far, can pertain to this situation as well (it can be reduced in particular to the same picture, at least symbolically, that we considered above in the context of the other version of discreteness, if we swap the 'colors' of points and blank spaces between the points, that is, if we draw white isolated points on a black piece of paper).

It appears that the notion of size of points (which in practice are always extended, as we have been arguing) matters only from the 'outside', that is, when different points are compared to one another (from the perspective of a 'bird's eye view'), but not from the 'inside' of a point when considered in isolation. This observation could be used and serve as a basis to posit a relativity of size (somewhat like described in \cite{Vassalloetal2022}). According to such a relational approach, the two configurations depicted in Figure \ref{Fig1}f (assuming for the moment that the gray points are not there) would be identical, and to traverse the distance between the top two points (moving with the same nominal velocity) would take the same amount of time in both the left and the right configuration, as everything in the right one is a scaled version of the left, including the gaps between the points as well as the definition of a "meter". This is so, however, only intrinsicaly for a given configuration, when it is considered in isolation from others. When the configurations are directly juxtaposed and compared, or when one is smoothly transitioning into another, then these two configurations are different, as the existing points need to be enlarged (in a process of some sort) or new points 'produced' and added to the left configuration (as depicted by the small gray points) in order to obtain the right one and so the considered distance turns into a relatively larger one (that is, larger when judging its size by counting the number of points of a given fixed size needed to fill it, which appears to be the common practice in considerations where such a relativity of size is not posited).

\section{Can two ideal point-like events happen neither at exactly the same moment of time nor at two distinct moments}

Is it possible, within the SF, to conceive of an example of ideal point-like events which happen neither at exactly the same moment of time nor at two distinct moments (in a given reference frame), thus defying in a way the law of excluded middle, as targeted in \cite{Jura2024}, while staying within the classical framework with ideal points? One could try to argue that it indeed is, taking an extended version of points\footnote{'Extended' numbers akin to this kind of extended points seem to appear in certain approaches, considered nonstandard, to analysis (discussed also in \cite{Lynds2003}), where they are meant to represent an increment (of a variable's value).} (as in the discrete models of 'physical continuum' as discussed above) to prove the point (Figure \ref{Fig1}g, left, with the time axis to be considered here as going horizontally). What can only be compared in the SF, however, are points situated along a given specific dimension. That is, what we would need to do in order to be able to compare the temporal (or any other\footnote{In fact the discussion in this paragraph could apply as well to, for example, spatial relations between events, if we assumed instead that the horizontal direction in Figure \ref{Fig1}g represents a spatial, rather than temporal, dimension.}) relations of such points in a fair and rigorous manner (being faithful to the SF), is to move the lower point upwards, so that they are both on (exactly) the same level. Then, however, they would merge and become one event, for which we assume it obviously makes no sense to compare its relations with itself (Figure \ref{Fig1}g, right). The only conclusion that can be drawn in the SF is that either some points are two distinct points or one and the same point. If one would like to insist (focusing on the left part of the Figure \ref{Fig1}g as it is) and argue that clearly, one end(-point) of the lower point falls within the interior of and its other end(-point) falls beyond the upper one, we would respond that, in fact, the exact 'end' of the object being represented by the extended point, as a point itself, is something which does not really exist, and it is not in any specific place (neither is any specific (sub-)point of the 'interior' of any of the extended points), and so it just does not make sense for such exact points that do not exist to be strictly compared. It is only an assumption of the SF that such specific positions can indeed be determined. One could argue here, however, that the same can be said about and applied to the situation when two such points are on the same level (even when merged, as in Figure \ref{Fig1}g, right), and then they also cannot be compared because they are not in any specific positions either. With such observation we would agree, and add that this is an artificial construct of the SF in which the only thing we can say, and are in fact obliged to decide on, is whether two objects are distinct and thus not the same object or, alternatively, the same object, and this requirement is (symbolically) represented strictly and fulfilled only when they are situated along one selected dimension (that is, having $x_1$ vs $x_2$). As discussed already above, we regard this kind of SF's 'extended' points, as considered here, equivalent to 'ordinary' points, assumed commonly to be non-extended (identified possibly with exact, specific locations in an infinitely divisible continuum), in that their essence is to represent distinct, separate objects. The point of our alternative (CC) view is rather that there are no such specific, exact points.

In general, in the SF, if there actually were to be no not-so-well-defined 'emptiness' between neighbouring points, the points would need to be packed so densely that there are no gaps between them (no 'white' space between 'black' points). Then, however, they would become simply one point.\footnote{To get rid of the emptiness, and arguably of any 'borders' between the points as well, while retaining distinct points, one could argue that it is the adjacent points themselves, constituting a continuous interval, that might be represented by the alternating black and white colors. We respond to this, hovewer, that the colors and the way we choose to color the points representing some concrete things can be arbitrary, and if the supposedly adjacent points were 'touching' tightly, then, again, they would in fact be one point. Undoubtedly oversimplifying greatly, we could also remark here that if there were to be regions in the natural world with different matter density, with matter represented by the 'static' points, without the emptiness, there would need to be points of somehow different kinds, with different intrinsic 'physical' properties, that is lighter or heavier (whatever that would mean). Obviously some, if not most, of these considerations and argumentations of ours concerning the SF resemble, or overlap with, some quite ancient ones, that can be found already in Aristotle when objecting to the notion of void separating the atomists' atoms, or Pythagoreans' numbers (of which they believed matter literally consits) \cite{Barbour2001}. We would like to avoid thus going even more deep into deliberations of these matters and reinventing the wheel by inadvertently repeating arguments that most likely can be found in some old texts, and instead just indicate that some of those arguments might as well apply to what is commonly considered as continuous intervals (as suggested, primarily in relation to the concept of time, also in \cite{Lynds2003}), and then focus on doing some constructive work (or, at least, suggest that such a work should, and perhaps can, be done), that is, departing from what we stated as our main postulate, try to see whether and how it can be developed into some intelligible structures or tools and whether it might consequently turn out fruitful in describing various observed phenomena.} It seems that there is just no third way within the SF and, because of this, oftentimes we are forced to deal with all "how many angels can dance on the head of a pin"-sort of 'problems'.

\section{Are real events spatio-temporally distributed and overlapping like fuzzy sets}

The CC viewpoint posists instead that events are not like ideal points, with their specific positions, but rather that their persistence can be considered graded, as represented with the use of fuzzy sets and ("spatiotemporally") blurred blobs \cite{Jura2024}. However, is it supposed to mean that those grades of events' persistence will take specific values at distinct points? Ref. \cite{Jura2024}'s Figure 1 might be misleading in that it suggests a picture of a function taking specific values at distinct specific points of a domain, that is, a specific value of the 'membership' function assigned to each point of the domain (constituted presumably by the points of a spacetime, that is, by 'underlying' events), with the 'strength' of an event's persistence being equal, for example, 0, 0.5, 0.99, or 1, as in Figure \ref{Fig1}h (which describes the situation equally well even if we assume that these are values of a 'density' function, as long as we insist on having specific values)\footnote{On such strict assignment of specific values appear to be based approaches in various areas of what is called fuzzy mathematics (deriving from \cite{Zadeh1965} and, more fundamentally, multi-valued logics).}. In contrast to that, the 'persistence' of events in our viewpoint is to be taken rather as having no specific value at any point. Consequently, one cannot take a strict cut-off over or below a given value, e.g., selecting points with $f(x) > 0.5$ (to determine the, approximate but nonetheless strict, extent of one event; and the very same argument would apply if one tried to supplement here the symbol '>' with '='). Also, it is not possible, for example, to take a strict intersection of two or more events, finding points where the corresponding functions would overlap (by taking the lowest value at each point), as there can be no two or more distinct objects, or functions (representing some actual objects) defined, at exactly the same point\footnote{As opposed to the SF where, typically, it is considered possible for two or more physical (althought to what extent they can be considered 'physical', instead of merely 'abstract', is not always, if ever, obvious) objects to be at exactly the same point (one specific example of a pair of such objects might be a spacetime metric and matter).} (due to there being, in the first place, no exact points strictly distinct from others, and thus no functions to be defined at such points).

\section{Do events occur next to each other in space and time}

What is to be conveyed by the use of fuzzy sets and the blobs, is rather that there is simply no clear-cut separation (of CC) into distinct events, that CC does not require from us deciding on whether it is one or multiple events, as it is better seen as a continuous 'interaction' (with the exact sense of this term as used in this context remaining to be specified), and even when it tends to 'separate out' what can be considered distinct events, they will still 'permeate' the rest, in a manner that can be considered graded.

Figure like the \ref{Fig1}e might suggest a picture where, according to the SF, events sit in a net of events, hold tightly and defined by (links with) other events. At the same time, if the gaps are taken to represent an empty space (whatever it would mean), it might equally well suggest a picture in which each event is detached from the rest and occurs fully independently. CC, in contrast, is to be seen as neither a rigid structure nor fully independent events separated and detached from the rest. There is no clear-cut separation ('spatial' or 'temporal', or otherwise) between events, they thus do not form a collection of neighbouring points being distinct along a specific dimension, with (more 'fine-grained') points where one would end or begin and with empty spaces between them (or, neither, with exact points of a 'weaker' persistence of events). The blobs depicted next to each other, as in ref. \cite{Jura2024}'s Figure 1, might be misleading in this regard (even if accompanied by a disclaimer saying that it is supposed to present an idealized situation, taken to the extreme, which, importantly, is qualitatively different from what we are trying to describe with this model), as putting them next to each other implies a specific dimension along which they would be distinct\footnote{Taking the example considered above based on Figure \ref{Fig1}c, left, it would be therefore not quite right if one tried to represent a process as described by the curve (or, alternatively, the process of constructing the curve per se), using for example the (rather abstract) angle, starting from the 'bottom' of the curve, as a series of blobs situated next to each other on an $angle$-axis (or $angle$-$time$-plane), possibly ordered by some increasing parameter, with a first blurred blob in the series 'centered' on an angle value representing moving in the vertical direction (but encompassing in a graded manner more than one value), the last blob centered on the horizontal direction, and the other ones centered on the intermediate values betwen the exactly vertical and the exactly horizontal ones, respectively, for such a separation into distinct blobs (as represented by the ordering parameter) would itself need to be considered graded, suggesting in our view that the very notion of a collection (of blobs situated next to each other) is simply not quite appropriate (and the same objection would apply if instead of 'angle', considered here abstract, we took an axis of (a spatial) '$x^1$', '$x^2$', ..., or any other relevant variable).}. CC instead is what can be considered a 'smooth' change.

\section{Where do events persist}

What can be considered spatial or temporal (or some other) separations likely emerge (in a graded manner) from CC, depending on events, and so events themselves do not 'occur', that is, persist and change, in space or time (in the sense of them being akin to points in, or \textit{of}, an abstract mathematical space, which could represent a space and time, or spacetime). For instance, if I went now to what I consider to be the office (or some other place) where the author of ref. \cite{Zadeh1965} was in what I consider to be a distant past writing his paper (or at least some part of it) I would not 'meet' him there, but this does not mean that the event of him writing the paper is not persisting (and, of course, changing) in a certain way\footnote{Perhaps like what we consider 'memories', which in such view could persist, but not in space or time (and also not 'arising' through simple 're-creations' of matter configurations similar to some past ones).}, in what can possibly be considered more of a 'present' by some other 'observers'.

\section{Do events proceed in a sequence}

Ref. \cite{Jura2024}'s Figure 1 might be misleading also in that it suggests a direction (by its use of the single-headed arrows), as if events emerging from CC were occurring in a sequence\footnote{Or -- assuming ideal SF-kind of events, if we wish to consider only a situation where identical events should not repeat, and in assessing whether two events are identical or not do not take into account their relation to other events, i.e., what we can consider their spatio-temporal location, but only intrinsic features, for instance a specific configuration of matter if for the moment by 'events' we understand such configurations -- as a directed set, rather than sequence.}. However, the primary focus of the CC viewpoint is on the separation process itself (as depicted with the double-headed arrow in Figure \ref{Fig1}i), likely being more essential. 'Spatial' or 'temporal' (or some other) nature of a separation between what is considered separate events, or direction of such separation, may be relative (dependent on a perspective, defined in some way), undefined, or possess yet some other characteristics. Our viewpoint remains rather agnostic and does not say anything specific about the direction or nature (or lack thereof) of such separation, that is besides its 'graded' and/or 'subjective' nature and its 'scale'.

Regarding the latter of these aspects, we can note that relatively more 'rapid' or 'slow' changes will determine different scales (as modeled in Figure \ref{Fig1}i by the length of the arrow, being proportional to the blob's 'diameter'). A slower change results in a 'stronger', more direct interaction with what is considered more 'distant' events, that is, distant when assessed using a scale determined by some relatively more rapid change. This does not mean, however, that what is considered a temporal history or space should be seen as composed of a multitude of small bits of rapid changes arranged next to each other (akin to the ideal points of SF), with bits of empty space in between, possibly imposed on a background constituted by a separate slower change (as could be suggested by ref. \cite{Jura2024}'s Figure 1e).

\section{Does everything happen all at once}

On the picture of neighbouring distinct points based is a notion of 'locality', in the sense of interaction only between neighbouring points, that is such situated directly next to each other (namely, with no other points in between), with the possible interactions between objects considered as not neighbouring termed action at a (potentially also temporal) distance \cite{Price2012,Adlam2018}. Such a distance can be considered a measure of numerosity of a collection of points of a given fixed size, as discussed already above, and thus our considerations pertain also to this notion of locality. The essential questions, in light of the above considerations, is how and to what extent events get separated, becoming distinct, and individuated (and remain so\footnote{And, then, how emerges what appears to be the kind of motion (or, in general, change) that is called loco-motion, \textit{motus localis}, in the sense of specific objects changing place over time \cite{Barbour2001}.}) \cite{DieksLubberdink2020}, and also, perhaps relatedly, how different events, or what is considered to be objects located in different places, can appear similar, that is, share certain properties, which allows to assign them to certain classes and describe their behavior (at least approximately) with the same laws. In this context, we note that CC is (inherently) a sort of interaction, and thus events need to be consistent, across both what is considered 'space' and 'time'. For instance, when I take a left turn (instead of a possible right) at a three-way corridor intersection, then in what I might consider the 'future' the world will necessarily (if only the events obey the relevant laws of physics) need to find itself in a situation with myself inside the left corridor. If at some point I find myself entering the left corridor at this intersection (arriving via the same route as in the example above), then in what I might consider the 'past' I have necessarily had to take the left turn. When I take the left turn, all events, however distant they could appear, need to agree and align with myself entering and occupying some space inside the left corridor (with all the events described in these examples to be understood in terms of CC). It does not mean, however, that everything occurs all at once, as in our viewpoint there is no notion of 'every-thing' (i.e., all the distinct things; unless in some meta-physical sense, which we cannot, or rather should not, rule out a priori \cite{HealeyGomes2022}), and also no notion of 'at once' (i.e., a single separate moment).

\section{Is the separation of events graded and/or subjective}

We considered in \cite{Jura2024} the events, as emerging from CC, to be related to, or dependent on, a subjective perspective (instead of the term 'subject', which we have tried to avoid thus far, we could adopt here the term 'observer', with the exact meaning of the term and process of 'observing', which is not necessarily to be a purely passive one, and in particular to what extent it overlaps with or replaces in such description the process of 'experiencing' that we have considered before, remaining to be specified), calling them 'neutrally' subjective, meaning that potentially everything can consitute an 'observer' of events, with the separation (or, detachment) of distinct events being itself a (potential) process that might be considered, at least to some extent, objective. How exactly, however, is this objectivity (and subjectivity) to be understood? In order to specify and clarify this point, we could ask the following questions: will other (potentially all) observers agree on whether a subjective event of one of the observers is occurring? would they agree on how exactly it is occurring (that is, how is it like, how does it 'look' like, metaphorically speaking)? According to the CC viewpoint, different observers will observe different events, 'filtering out' and 'integrating' different 'portions' of CC, depending on the observer including its history and thus structure, with the resulting events being to some extent 'detached' from the rest, and thus others cannot know exactly how a subjective event of a given observer is like. However, if the separation is graded (in certain sense), then each such event still 'interacts' with and to some degree affects all the rest, and hence other observers can always in principle know \textit{that} it (or, rather, something) is occurring (but not \textit{when} or \textit{where} exactly it is occurring).

Is the separation into distinct events indeed graded, that is, stronger or weaker (and, then, how is this 'strength' to be understood)? If so, is it graded objectively, that is, will other observers agree that a given event (or, at least, some, not necessarily \textit{the same} according to different observers, events) are more or less separated and distinct from the rest? Can it be graded subjectively, that is, with an observer considering a given event as more or less separated, independently from the other observers' 'opinion' on this matter? Does it then correspond to a more 'intrinsic' or 'extrinsic' mode of 'observation' of events? Can it involve 'illusions' of some sort? Also, how, and in what sense, would such more or less separated events 'persist' (and 'change')? In any case, without trying to answer these complex questions here, such a degree of separation of events could be modeled adjusting the relative color intensities of the double-headed arrow (with heads pointing in opposite directions, symbolizing the separation process), and the loop, as in Figure \ref{Fig1}i, bottom (the color intensities of the tangent circles relative to each other are determined by the blob's diameter, that is, its scale, and so all should change roughly proportionally).

\section{Discussion}

\subsection{Does negating the static separatedness of points lead uniquely to the dynamic CC}

Does negating the 'static' picture of the SF imply the 'dynamic' CC\footnote{'Dynamic' ('static') typically means something that does (not) change over time (or, in general, at transitions from one state to another). For discussions on whether an event (or time) in itself, when there is no independent passage of time, can be considered 'dynamic' (or likewise 'static') see \cite{Jura2022,Lynds2003}. In short, the solution to this problem is in our view simply to find some better term that will work for the reader (our take: '(not) continuously changing'). The point is that CC is simply something different from the familiar picture of states situated next to each other (possibly at distinct instants of time).}? Is it a unique alternative, or are there others? Rather than a simple, exact negation, we see CC as a generalization, with the separation of distinct points, which would neither change nor persist, not 'interacting' with the rest, being certain approximation (never realized exactly) and perhaps only a small subset of the domain that we are trying here to find a way of thinking about and describing. How does this approach of ours relate to others, in particular such concerned with the very concept of 'point'?

\subsection{Are fuzzy or point-free frameworks free from ideal points-like separatedness}

To our knowledge and understanding, different areas of what is called fuzzy mathematics, as well as different point-free frameworks\footnote{Like geometry \cite{Gerla1995}, or topology \cite{Johnstone1983}, doing without the notion of point (at least as a primitive, primary one).} (with various motivations behind their development, and being studied from different perspectives), from our perspective and for our purposes do not go far enough in this regard and are not free from the property of separatedness (which always sneaks in one way or another, as soon as we try to conduct any practical reasoning). What some of these approaches (like the ones described in \cite{Gerla1995}) have in a sense in common with ours (regardless of the rather specific, general definition of 'point' that we employ here), is that they try to derive the notion of point from some other, more basic ones (in our case arguably also a more intuitive).

\subsection{Can we possibly conceive an accurate mathematical model of the physical world that would not contradict our direct experience of this world}

One could argue that it is not well justified (in light of lessons from the history of science) to expect from a serious model, or theory, aspiring to be accurately describing the natural world to agree with our intuitions based on our direct experience of this world. After all, one could point out, for example, that the heliocentric model, undoubtedly more convenient and thus appropriate than its direct rival, the geocentric one, does not agree with our perceptions and experience of the world, since it contradicts the (common) observations of the sun raising in the east in the mornings and then moving along the immobile Earth. We can use this particular example, however, to further clarify what is supposed to be meant by 'direct' experience. To this end, we would argue that no one, including people unaware of the heliocentric model, has ever experienced directly the planet Earth at rest at the center of the cosmos (or any system) with sun revolving around it (as was posited by the geocentric model). Rather, what can only be observed directly, is the sun moving continuously relative to Earth (e.g., falling into the horizon) \cite{Barbour2001}. In other words, models like both of these two, even the one considered more 'intuitive', are essentially abstract models, constructed by fitting certain direct observations into the SF and thus describing them by abstract concepts not easily reconciliable with our direct experience.

More adequate approach would be to try to find means of description of the world such that, in the first place, our direct experience of the world so described can be the way it is. The essential aspect of this experience is, in our view, what we might consider its 'flow' (taken to be an actual change), capturing of which is the main (positive) ambition of CC. Does the CC viewpoint, however, besides potential realization of this aspiration, allow also to capture and describe (at least certain features of) the empirical observations and findings obtained in other, more indirect ways, and make some new testable predictions, in particular predictions (or explanations) of shortcomings of certain current methods and theories (perhaps related to their inadequacy stemming from them being used to domains beyond the scope of their applicability)?

\subsection{Could we measure values of physical quantities with arbitrary precision}

One of predictions (or, explanations) suggested by the CC viewpoint is that models based on the SF's notion of ideal point, trying always to determine and 'pinpoint' specific separate points in actual phenomena (that is, specific values of quantities describing a physical system), will fail when encounter the actual continuity, the 'smooth' change, of CC instead. This will happen more readily when dealing with what can be considered relatively small, rapidly changing events, as for us, from our perspective, they can be 'seen' directly and thus studied more easily, appearing as if they indeed constituted separate events, or objects (as seen under the microscope). This part of the 'prediction' appears to be, in fact, accurate, as suggested by empirical findings \cite{GriffithsandSchroeter2018}. However, the same prediction pertains also (as suggested also in \cite{Lynds2003}) to models of this kind when applied to relatively large, slowly changing events, and their (potential) attempts of representing and describing such events with the use of ideal points and treating thus as separate entities, for instance, a 'chair', 'table', or 'planet' (and perhaps what could be considered a whole 'space' itself). Why, then, one could object, such objects seem to be spatially and temporally relatively precisely located and separate? In our view this is because they do not change (relatively) rapidly, that is, they are more 'blurred', 'integrating' over larger ('time' and 'space') intervals. A chair is something that we do not see directly. What we can see more readily are what can be considered parts (more rapidly changing) of which it is composed. That something cannot be seen 'under the microscope', however, does not necessarily mean that it does not exist. We can experience the undoubted existence of some chair by, for instance, sitting on it.

The proper domain and scope of applicability of methods (like calculus \cite{calc,int}), models, or rules and laws, based on the SF's notion of ideal point, is simply limited. We have tried here to clarify why we think this is so and specify what is its extent and what might lie beyond it. In particular, since laws based on the SF are ultimately an abstraction \cite{Smolin2015}, they will change with events. The interesting questions we could ask in this context are: how, and how 'rapidly'.

\subsection{How rapidly are specific laws changing}

In order to address the latter of these two questions, and establish the bounds on the rates of change of specific laws, one could compare the results of experiments considered analogous performed at different points in time or space (in a given reference frame), taking into consideration accuracy with which the measurements were made, as well as the size of the relevant interval between consequtive measurements. In principle, relatively those rates might seem to be 'infinitely' low, with the laws appearing thus as non-changing. In practice, however, one should rather expect that they will be bounded from below (in a \textit{graded} manner) by some 'slowest' CC. But what does our viewpoint have to say, one could ask, about the content of the particular laws?

\subsection{Why do we observe and remember only the kinds of events that we do and not others}

Why do we observe, and remember, only certain kinds of events and not others? For instance, we can observe a glass breaking into pieces when dropped on the ground from a sufficient height, but typically we do not observe the pieces of a shattered glass getting spontaneously, by themselves, put back together? In our view, it is simply because the observed phenomena follow the laws of physics. In this particular case, the relevant laws happen to say that it is possible for a glass to break into pieces when dropped on the floor from a considerable height, and that it is highly unlikely (if not impossible) for pieces of broken glass to re-arrange spontaneously. We may however eventually observe such an unlikely process (but perhaps not anytime soon), if only it is permitted by relevant laws. The CC viewpoint remains rather salient and does not say anything specific about the content of such specific laws in this regard. What it does say, however, is that when such a potential phenomenon actually occurs, that is, pieces of broken glass getting put back together spontaneously (or, similarly, a glass breaking down), it will always be a CC, which does not have an inherent direction, but which persists (and changes) and thus can be considered 'irreversible' (and 'indeterministic', as we noted already in \cite{Jura2024}).

\subsection{How indeterministic might indeterminism be}

We pointed out and discussed here the reasons why certain aspects of our viewpoint might be ambiguous and potentially misleading, and did so, rather than specifying what exactly it is, primarily by trying to clarify what it is not, how it should not be interpreted, analyzing its relation and, for the most part, putting it in opposition to various notions used commonly as bases of models with ambitions of describing accurately certain physical systems and natural phenomena in general. This mostly negative approach stems from our conviction that the potential usefulness of the viewpoint considered here lies primarily in indicating what are the proper questions to ask, rather than giving definite specific answers. This view, in turn, is based in particular on the conclusion that as an essential element this viewpoint posits certain form of what we can consider an 'indeterminism'. First of all, thus, we should clarify how indeterministic this indeterminism might be, that is, what exactly might be its nature and its extent. As such, in itself, a model of this kind will not have any immediate practical applications, but it might be useful indirectly, by potentially sparing certain misdirected efforts, informing various attempts of constructing specific physical theories about those of their aspects that are particularly questionable, suggesting what is most likely futile and what not to expect from the natural phenomena, and indicating more promising directions of inquiry. It suggests, in particular, not to expect an actual existence (i.e., in the sense of 'ontology') of things like: strictly separate systems, specific instants of interactions between such separate systems, specific states of systems, different exact probabilities of such states being observed by an observer, deterministic evolution or transitions between such states, or specific events (or, values of physical quantities) observed by an observer (nor possible alternative events observed by other, 'branching-off', strictly distinct observers).

\section{Conclusion}

Is the natural world static and immobile or rather in a state of constant flux. Does it consist of discrete bits of matter floating suspended in a void, or of continuous fields. From objects, or processes. Beings, or becomings. Is it rather discrete, or continuous. All those views (with some of them, sometimes, but not always, considered as opposite of each other), if formulated in terms of the ideal mathematical points, being fit into the SF, have a common underlying assumption which we see as a not necessarily well-justified one, and, let us conclude, it is simply good, on many levels, to be aware of it.

\end{multicols}
\end{document}